\documentclass{article}
\usepackage{mrs2005,epsfig}
\begin{document} 
\title{THE CORRELATION OF THE LYMAN-$\alpha$ FOREST AND METALS IN CLOSE PAIRS OF HIGHT-REDSHIFT QUASARS}

\author{F. Coppolani$^1$, P.~Petitjean$^2$,$^3$, 
F. Stoehr$^2$, E.~Rollinde$^2$,
C. Pichon$^2$, S. Colombi$^2$,M. Haehnelt$^4$, B. Carswell$^4$, R. Teyssier$^2$}
\affil{
\begin{tiny}
$^1$ European Southern Observatory, Alonso de C\'ordova 3107, Casilla 19001, Vitacura, Santiago, Chile 
 - email: fcoppola@eso.org \\
$^2$ Institut d'Astrophysique de Paris, UMR 7095 CNRS \& Universit\'e Pierre et Marie
Curie, 98 bis boulevard d'Arago, 75014 Paris, France \\
$^3$ LERMA, Observatoire de Paris, 61, avenue de l'observatoire F-75014
        Paris, France \\
$^4$ Institute of Astronomy, Madingley Road, Cambridge CB3 0HA, UK\\
\end{tiny}
}

\begin{abstract} 
We derive the transverse flux correlation function in the 
Lyman-$\alpha$ forest at $z$~$\sim$~2.1 from VLT-FORS observations
of a total of 32 pairs of quasars. The shape and correlation length of the transverse correlation 
function are in good agreement with the paradigm of intergalactic medium 
predicted in CDM-like models for structures formation. 
Using a sample of 139 C~{\sc iv} systems detected along the lines of sight 
toward the pairs of quasars we investigate the transverse correlation of metals
on the same scales. 
We find that the correlation function is consistent
with that of a randomly distributed population of C~{\sc iv} systems. 
However, we detect an important 
overdensity of systems in front of a quartet.
\end{abstract} 
 
\section{Introduction} 
 The  intergalactic medium (IGM) is  revealed  by  the numerous  H~{\sc  i}
absorption  lines seen  in the  spectra  of distant  quasars, the  so-called
Lyman-$\alpha$ forest. There is a long history of using the 
flux transverse correlation in the Lyman-$\alpha$ forests of QSO pairs to  
measure the spatial extent of the corresponding absorbing
structures. The Lyman-$\alpha$  forests in the spectra of multiple 
images of lensed quasars or pairs of quasars with  separations of the order of a few
arcsec (Smette et al. 1995; Impey et al. 1996) 
appear nearly identical implying that the absorbing 
structures have sizes $>$50$h_{75}^{-1}$~kpc. Significant correlation 
between absorption spectra of adjacent lines of sight toward quasars 
still exists for separations of a few to ten arcmin suggesting 
dimensions or better a coherence length of the structures larger than
500~$h_{75}^{-1}$~kpc (e.g. Petitjean et al. 1998; Aracil et al. 2002).

\section{The transverse and longitudinal flux correlation functions}

We derive the transverse and longitudinal flux correlation
functions from 32 QSO pairs. 
We use $N$-body simulations to understand the effect of thermal broadening and 
peculiar velocities, evaluate uncertainties and interprete the observations.   
The comparison between the observed and predicted transverse and
longitudinal correlation functions 
is a valuable test of the theory of the Intergalactic Medium. 
In addition, by fitting the two correlation functions it may be possible, 
as a final goal, to constrain the cosmological geometry through the 
Alcock \& Paczy\'nski (1979) test 
as proposed by McDonald \& Miralda-Escud\'e (1999) and Hui et al. (1999)
(see also Rollinde et al. 2003). This goal will be achieved only with
a major observational effort. Indeed, it has been shown that this
should require a number of pairs as large as 13($\theta$/1')$^2$ on scales
at least up to 10~arcmin (McDonald 2003). It is therefore crucial as a 
first step and before embarquing in a large survey to demonstrate that 
the general theoretical scheme is consistent with observations.
\par\noindent
Each quasar pair yields one data point of the transverse correlation function.
The observed transverse correlation coefficients measured from the 32 pairs are 
shown in Fig. 1. 
We have gathered the observed measurements in bins weighting the points by the errors. 
Overplotted as a continuous line is the correlation function measured in the 
hydrodynamic simulation at $z$~=~2. 
\par\noindent
Predictions from linear theory are compared for different values of $\Omega_{\rm m}$.
Despite the larger sample than 
in Rollinde et al. (2003), we cannot distinguish between different values of $\Omega_{\rm m}$. 
However, our result demonstrates the presence of a strong signal
in the transverse correlation on scales $<5$~arcmin and justifies 
the investment in observing time that will be needed to extract
the information on $\Omega_{\Lambda}$.
\par\noindent
It can be seen that the shape of the observed  function is consistent with what is
expected both from the simulations and the linear theory.
Besides, if one assumes the favored LCDM cosmology, with $\Omega_{\rm m}$~=~0.3, 
a non linear signal is probably present in the transverse direction 
for $\theta<1$ arcmin, as expected from the fully non-linear hydrodynamic results.  
This demonstrates that data are consistent with the current scheme describing the 
Lyman-$\alpha$ forest as the tracer of a continuous photoionized and warm 
intergalactic medium as associated with the filamentary and sheet-like
structures predicted in cold dark matter-like models for structure formation.
\begin{figure}  
\begin{center}
\epsfig{figure=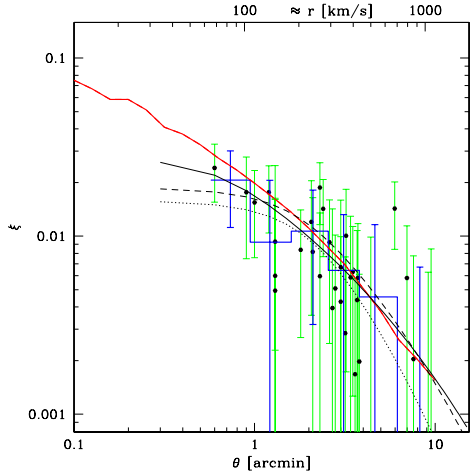,width=6.5cm}  
\end{center}
\caption{The observed transverse correlation coefficients
for individual pairs (black dots) and binned 
(histogram) are overplotted over the transverse correlation
function from the hydro-simulation
and  linear predictions for $\Omega_{\rm m}$~=~0.1, 0.3 and 1 
(thin solid, dashed and dotted lines respectively)
} 
\label{f:comparison} 
\end{figure}

\section{C~{\sc iv} systems}

To estimate the correlation of intervening C~{\sc iv}~ systems in the QSO pairs, 
we apply the Nearest-Neighbor method , as described in Young 
et al. (2001) and Aracil et al. (2002) to a list of selected C~{\sc iv}~ systems.
 For each absorption line along one QSO line of sight, 
we search the adjacent QSO line of sight for the nearest (in velocity) 
absorption line.
\par\noindent
There is a possible small excess of clustering of C~{\sc iv} systems on scale smaller
than 5000~km~s$^{-1}$. 
This scale is larger than the typical correlation length,
about 1000~km~s$^{-1}$, seen in the longitudinal correlation function of
C~{\sc iv} systems (Rauch et al. 1996, Pichon et al. 2003,
Scannapieco et al. 2005). 
This is due to an excess of associations in the bin $\Delta V$~$\sim$~4000~km~s$^{-1}$.
Looking in more details at the lines of sight from which this excess
comes from shows that most of the corresponding C~{\sc iv} associations 
are located in 
the peculiar field of the quartet Q~0103$-$295A\&B, Q~0102$-$2931 and Q~0102$-$293,
 we find 9 out of the 25 C~{\sc iv} with $\Delta V$~$<$~5000~km~s$^{-1}$.

\section{Conclusion}
The shape and correlation length of the observed transverse correlation 
function are in good agreement with those expected from absorption by the 
filamentary and sheet-like structures in the photoionized warm intergalactic medium 
predicted in CDM-like models for structures formation. 
However, it is apparent from Fig.~\ref{f:comparison} that although the sample
is unique, it is not sufficient to significantly constrain
the geometry of the universe. This confirms the predictions by
McDonald (2003) that significant constraints can be obtained on 
$\Omega_{\Lambda}$ only with a large number of pairs of the order of
13($\theta$/1')$^2$ on scales at least up to 10~arcmin.
Our result demontrates the presence of a strong signal
in the transverse correlation on scales $<5$~arcmin and justifies 
the investment in observing time that will be needed to extract
the information on $\Omega_{\Lambda}$.
\par\noindent
No correlation signal is seen in metal lines (C~{\sc iv})
on the scales probed by our pairs. This is not surprizing as most
of the separations are larger than 2 comoving Mpc derived by Scannapieco et al. (2005)
for metal-enriched bubbles surronding massive haloes.
Finally, we detect an important 
overdensity of systems in front of the quartet Q~0103$-$295A\&B, Q~0102$-$2931 
and Q~0102$-$293 extended over the redshift range 1.7~$\leq$~$z$~$\leq$~2.2 and on
a spatial scale larger than 10~arcmin.



\vfill 
\end{document}